\documentclass[%
 aip,
 apl,%
 reprint,%
]{revtex4-1}

\usepackage{graphicx}
\usepackage{dcolumn}
\usepackage[utf8]{inputenc}
\usepackage{bm}
\usepackage{amsmath}
\usepackage[x11names,table,xcdraw]{xcolor}
\usepackage{subcaption}

\usepackage{soul}

\begin{document}

\title[Spatial emerging texture in simulated memristive arrays]{Spatio-temporal evolution of resistance state in simulated memristive networks}

\author{F. Di Francesco}
\affiliation{ECyT-UNSAM, Martín de Irigoyen 3100, B1650JKA, San Martín, Bs As, Argentina.}

\author{G. A. Sanca}
\affiliation{ECyT-UNSAM, Martín de Irigoyen 3100, B1650JKA, San Martín, Bs As, Argentina.}

\author{C.P. Quinteros}
\email[Corresponding author: ]{cquinteros@unsam.edu.ar}
\affiliation{ECyT-UNSAM, CONICET, Martín de Irigoyen 3100, B1650JKA, San Martín, Bs As, Argentina.}

\date{\today}

\begin{abstract}
Originally studied for their suitability to store information compactly, memristive networks are now being analysed as implementations of neuromorphic circuits. An extremely high number of elements is thus mandatory. To surpass the limited achievable connectivity - due to the featuring size - exploiting self-assemblies has been proposed as an alternative, in turn posing more challenges. In an attempt for offering insight on what to expect when characterizing the collective electrical response of switching assemblies, in this work, networks of memristive elements are simulated. Collective electrical behaviour and maps of resistance states are characterized upon different electrical stimuli. By comparing the response of homogeneous and heterogeneous networks, we delineate differences that might be experimentally observed when the number of memristive units is scaled up and disorder arises as an inevitable feature.  
\end{abstract}
   
\maketitle

\noindent Assemblies of memristive devices are attracting more attention due to their versatility to implement a variety of promising technological implementations\cite{chua_handbook_2019}. Originally claimed to be an alternative to efficiently increase information storage\cite{li_memristive_nodate,zahoor_resistive_2020} 
they are now at the focus of an intense research effort to develop neuromorphic circuits\cite{hochstetter_avalanches_2021}. Since memristive properties have been assimilated to particular synapses' and neurons' electrical responses\cite{sah_brains_2019,chua_memristor_2019,cai_synapse_2019} assembling memristive units into a statistically representative number appears as an inevitable strategy to compare them to their biological counterparts. Moreover, it has been proposed that a large enough collection of synaptic weights could demonstrate emergent behaviour\cite{chialvo_emergent_2010} capable of partially project in hardware operations that, otherwise performed in software, are complex and inefficient\cite{wang_deep_2019,karbachevsky_early-stage_2021}. The key to unveiling these phenomena is to increase the amount of units in the system.   

One possible approach is to organize memristive units in a pre-defined manner and scale them up as much as possible. An alternative strategy claims that the only way to achieve a statistically representative number of connections (with certain resemblance to biological brains) is to promote self-assembly of affine systems\cite{chialvo_emergent_2010}. In the latter, the number of units increases so dramatically that the challenge becomes how to perform useful operations with them. In this work, assemblies of memristive models are simulated to explore the collective electrical response and the lateral distribution of resistance states demonstrated by the formed network. The aim is to offer insight on how the networks collectively react and which type of variation is expected in the overall electrical response upon changes of different nature.     

A memristive device is an already old concept that has been around, at least theoretically, for the last 50 years\cite{chua_memristor-missing_1971}. Although its experimental realizations have been ubiquitous for many decades, a device was identified as such slightly longer than a decade ago\cite{strukov_missing_2008}. Even though its precise definition - and even the need for considering it as one fundamental circuit element\cite{abraham_case_2018,pershin_memory_2011} - is still disputable, the term is used to refer to a unit able to show different conducting states tunable by external stimuli. Its associated mathematical equations\cite{pershin_memory_2011} are

\vspace{-0.5cm}

\begin{equation}
I (t) = R_M^{-1}(X, V_M, t) \cdot V_M(t) 
     \label{eq:mem-gen-Ohm}
\end{equation}

\vspace{-0.8cm}

\begin{equation}
\frac{dX}{dt} = f(X, V_M, t) 
     \label{eq:dX-dt-gen}
\end{equation}

\vspace{-0.1cm}

\noindent where eq. \ref{eq:mem-gen-Ohm} is the generalized Ohm equation for the memristance (acronym for \textbf{memr}istive res\textbf{istance}) $R_M$ which, in turn, is a function of the internal state $X$, the time $t$, and the voltage applied to the device $V_M$. In memristive devices, $X$ is the internal state accounting for the changes to which the device has been subjected (eq. \ref{eq:dX-dt-gen})\cite{pershin_memory_2011}.

While eqs. \ref{eq:mem-gen-Ohm} and \ref{eq:dX-dt-gen} comprise the general expression, specifying $R_M(X, V_M, t)$ and $f(X, V_M, t)$ defines a specific model. We have chosen the one by Pershin and Ventra\cite{pershin_spice_2013} which allows to mimic both threshold-like and progressive switching types upon parameters modification. The model is as follows

\vspace{-0.5cm}

\begin{equation}
R_M \equiv X
     \label{eq:mem-Pershin-Rm}
\end{equation}

\vspace{-1cm}

\begin{multline}
f(X, V_M, t) = \\ f(V_M) \cdot [\theta(V_M) \cdot \theta(R_{OFF} - X) + \theta(-V_M) \cdot \theta(X - R_{ON})] 
     \label{eq:dX-dt-Pershin-f}
\end{multline}

\vspace{-0.6cm}

\begin{equation}
f(V_M) = \beta \cdot [V_M - 0.5 \cdot (\lvert V_M + V_t \rvert - \lvert V_M - V_t \rvert)] 
     \label{eq:dX-dt-Pershin-alpha0}
\end{equation}

\noindent where $X$ is assimilated to $R_M$ (eq. \ref{eq:mem-Pershin-Rm}) and $f(X, V_M, t)$ is reduced to $f(V_M)$ weighted by Heaviside functions (denoted as $\theta$) to restrict the resistance values (between $R_{OFF}$ and $R_{ON}$) and define a bipolar device (eq. \ref{eq:dX-dt-Pershin-f}). $f(V_M)$ is a null function until $\vert V_M \vert = V_t$ is satisfied determining the onset of a linear function with a rate of change $\beta > 0$ (eq. \ref{eq:dX-dt-Pershin-alpha0}). $f(V_M)$ is positive (resistance increase or RESET) for $V_M > V_t > 0$ and negative (resistance decrease or SET) for $V_M < -V_t < 0$. The model parameter $\beta$ sets the rate of change of $X$ ($=R_M$) in turn determining the type of switching. 

Each memristive unit is computed using NGSPICE\cite{noauthor_ngspice_nodate} by means of a sub-circuit whose internal state ($X$) is governed by eq. \ref{eq:dX-dt-Pershin-f} inserted in eq. \ref{eq:dX-dt-gen}. This sub-circuit is composed of a variable resistor, a linear voltage-controlled current source, 
and a capacitor\cite{pershin_spice_2013}. Additionally, we have included a modification suggested by Vourkas and Sirakoulis\cite{vourkas_memristor-based_2015} to effectively limit the resistance states between the programmed boundaries ($R_{OFF}$ and $R_{ON}$) by means of including two voltage sources mediated by opposed connected diodes. Complementary, NGSPICE's batch mode enables to define the initial state of each unit ($R_{init}$), replicate them as much as desired, and finally connect them to build the network.  

In this study, the externally applied voltage ($V_{ext}$) is always sinusoidal, with variable amplitude ($A$), frequency ($f$), and cycle repetition ($cyc$).  

\begin{figure}[ht!]
    \includegraphics[width=8 cm]{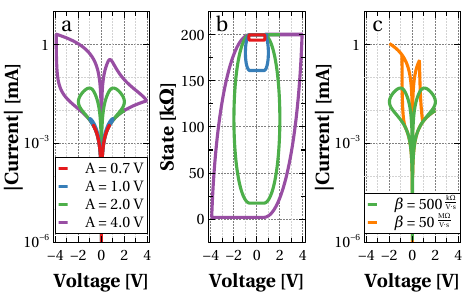}
    \caption{Memristive response of isolated units, defined as $R_{init} = R_{OFF} = 200$ k$\Omega$, $R_{ON} = 2$ k$\Omega$, $V_t = 0.6$ V, upon one sinusoidal cycle of $1$ Hz. (a) Current as a function of voltage (I-V) using fixed $\beta$ ($= 5 \cdot 10^5~\frac{\Omega}{V \cdot s}$) under four different $A$ values. (b) $X$ as a function of $V_{ext}(t)$ corresponding to the same runs shown in (a). (c) I-V obtained upon fixed $A$ ($= 2.0$~V) for two different $\beta$ values. }
 \label{fig:Ind-device} 
\end{figure}

\noindent Fig. \ref{fig:Ind-device} displays the response of an isolated memristive device ($R_{init} = R_{OFF} = 200$ k$\Omega$, $R_{ON} = 2$ k$\Omega$, $V_t = 0.6$ V) upon one sinusoidal cycle of $1$ Hz. Fig. \ref{fig:Ind-device}a presents current as a function of voltage (I-V) for a fixed $\beta$ under four different amplitudes $A$ of the externally applied signal. The obtained behaviour exemplifies what was mentioned before regarding the ability of Pershin's model to reflect multiple memristive types. As indicated by eq. \ref{eq:dX-dt-Pershin-f}, once the condition $V_{ext} = V_t$ is reached, $X$ starts varying. However, the effective ratio ($r_{eff} = \frac{R_{high}}{R_{low}}$) depends on how fast the switch is allowed to be ($\beta$) and how many steps the simulation performs before reaching the programmed amplitude ($A$) of the externally sourced signal ($V_{ext}$\footnote{$V_M$ and $V_{ext}$ are equal only when isolated memristive units are considered. In networks, $V_{ext}$ is distributed among the multiple elements forming the assembly.}). For that reason, even though $A$ is always higher than the chosen $V_t$, the selected $\beta$ value determines a small increment that becomes a complete switch only for $A = 4.0$ V (Fig. \ref{fig:Ind-device}a). Fig. \ref{fig:Ind-device}b clearly reflects this situation by quantifying $X$ as a function of the $V_{ext}$. A progressive-like switching is thus obtained for such $\beta$ choice allowing to appreciate minor loops, in which $r_{eff} = \frac{R_{high}}{R_{low}} < r = \frac{R_{OFF}}{R_{ON}}$. Oppositely, Fig. \ref{fig:Ind-device}c shows that the threshold-like switching, as originally proposed by Pershin and Ventra\cite{pershin_spice_2013}, is recovered upon an appropriate $\beta$ choice\footnote{It is worth mentioning the $\beta$ choice also depends on the particular combination of $A$, $R_{OFF}$, and $R_{ON}$}. 

Assembling memristive devices necessarily rely upon a geometry definition. Although our final goal is to tackle the completely disordered and spontaneous case of self-assemblies (such as Ag nanowires\cite{hochstetter_avalanches_2021,diaz-alvarez_emergent_2019,zhu_information_2021} or conducting ferroic domain walls\cite{catalan_domain_2012,seidel_conduction_2009}) here we present results indicating that even in the ordered case a hierarchy exists conditioning what material scientists might expect when intentionally introducing changes in experimental systems.    

\begin{figure}[ht!]
    \centering
    \includegraphics[width=8 cm]{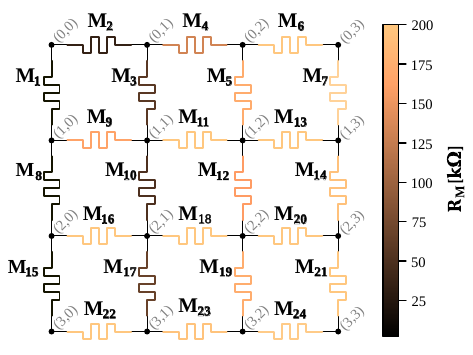}
    \caption{Resistance map of a memristive network comprising 4x4 nodes and 24 memristive units collapsing every device's $X$ in a coloured scale.}
    \label{fig:array-resistive-map}
\end{figure}

\noindent Memristive networks are built using a custom script in Python3 which generates a netlist that will be fed to the NGSPICE engine. The networks consist of a collection of nodes linked by edges. They are initialized by defining a geometrical array of square-distributed nodes (NxN) placing one memristive device in each horizontal or vertical edge. Fig. \ref{fig:array-resistive-map} shows the square-shaped 4x4 network which serves as the platform for the coming simulations. It contains 24 memristive units, each of them having a defined polarity (since the implemented model is bipolar). Unless stated otherwise, the network is homogeneously initialized both regarding the model parameters (coincident for every unit until further notice) and its polarity (with the cathode pointing down or right depending on the edge's orientation). The script allows for further distortions (such as memristive model exchange, nodes removal, units' reversal, and other means of introducing heterogeneity) that are not explored in this communication for the sake of length. 

\begin{figure}[ht!]
    \centering
    \includegraphics[width=8 cm]{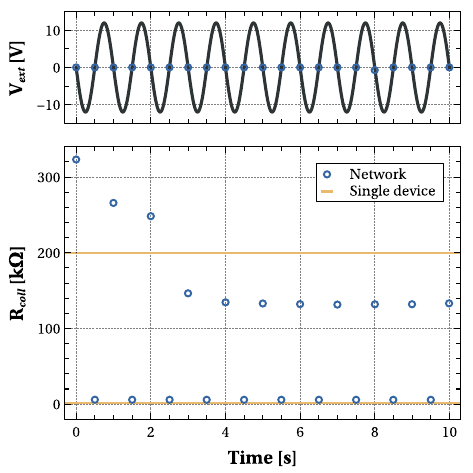}
    \caption{Collective resistance state (\textbf{R$_{coll}$}) as a function of time (lower panel) of an homogeneous 4x4 network ($R_{OFF} = 200$ k$\Omega = R_{init}$, $R_{ON} = 2$ k$\Omega$, $\beta = 500$ $\frac{k\Omega}{V \cdot s}$, $V_t = 0.6$ V) upon ten sinusoidal cycles of $1$ Hz with $A = 12$ V (upper panel). The resistance states associated to a single device are included for reference. }
    \label{fig:global-resistive-state}
\end{figure}

\noindent In the following, a set of simulations upon different externally applied stimuli is performed. The signal is sourced between the nodes (0,0) and (3,0) (see labels in Fig. \ref{fig:array-resistive-map}), with the ground in the latter. The network's electrical response is quantified as a collective resistance (\textbf{R$_{coll}$}) obtained as the linear fit of the applied voltage ($V_{ext}$) as a function of the current flow between (0,0) and (3,0), determined for each remnant condition (vicinity of $V_{ext}=0$ V). Complementary, resistance maps are included in the Supplementary Material to help visualizing the spatial evolution of every unit's internal state.

The results of a cycling experiment using an homogeneous 4x4 network are presented in Fig. \ref{fig:global-resistive-state}. Every device is set as $R_{init} = R_{OFF} = 200$ k$\Omega$, $R_{ON} = 2$ k$\Omega$, $V_t = 0.6$ V, $\beta = 500$ $\frac{k\Omega}{s \cdot V}$ while the external signal $V_{ext}$ is defined as depicted in the upper panel ($A = 12$ V, $f = 1$ Hz, $cyc = 10$). Fig. \ref{fig:global-resistive-state}'s lower panel demonstrates the evolution of \textbf{R$_{coll}$} upon the applied signal, calculated at each remnant condition (identified as circle symbols in Fig. \ref{fig:global-resistive-state}'s upper panel). The response of an isolated memristive unit upon the same external input is included for comparison. As it can be observed, initializing every device as $R_{OFF}$ determines \textbf{R}$_{coll}(t=0 ~s) \sim 330$ k$\Omega$. From that moment on, negative and positive sinusoidal semicycles are alternated, switching the network between high (\textbf{R$_{coll}^{high}$}) and low (\textbf{R$_{coll}^{low}$}) resistance states. Nevertheless, the specific values associated to them (mainly the high resistance ones although changes might be also occurring at the low range without getting noticed) vary upon cycling. After a transient of such behaviour, the network reaches a sustainable switching between two values, namely \textbf{R$_{coll}(t = 4 ~s)$} and \textbf{R$_{coll}(t = 4.5 ~s)$}. Compared to the case of an isolated device (included as lines in Fig. \ref{fig:global-resistive-state} for visual reference), the network displays a richer dynamics based on the underlying evolution of its components (see resistance maps at the Supplementary Material).   

\begin{figure} 
    \includegraphics[width=8 cm]{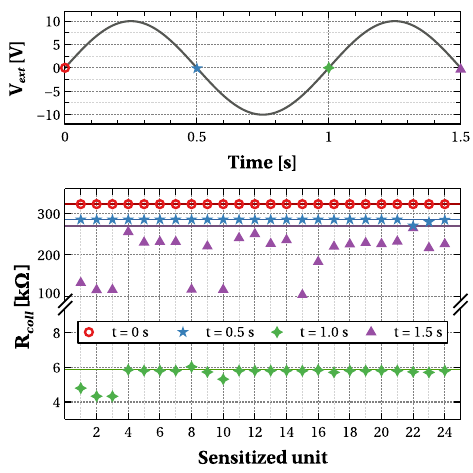} 
    \caption{Sensitization experiments. Upper panel: $V_{ext} (t)$. Lower panel: \textbf{R$_{coll}$} as a function of the position of the sensitized unit. The four different symbols (color online) account for \textbf{R$_{coll}$} quantified at the subsequent remnant conditions (consistently depicted in the upper panel).}   
 \label{fig:Sensitization-experiment} 
\end{figure} 

We then proceed with a sensitization experiment. Instead of programming each unit homogeneously, $V_t$ is set the same for every device except for one whose threshold ($V_t^S$) is dramatically reduced (sensitized). Since we want to explore the impact of this change depending on the location of the sensitized unit, 24 experiments are run. Each run corresponds to sensitizing a different unit whose position is raster within the whole network. Configuring $V_t = 0.6$~V and $V_t^S = 0.06$ V, $R_{init} = R_{OFF} = 200$ k$\Omega$, $R_{ON} = 2$ k$\Omega$, $\beta = 500$ $\frac{k\Omega}{V \cdot s}$ and applying a $10$ V-high $1$ Hz-sinusoidal signal, Fig. \ref{fig:Sensitization-experiment} shows the evolution of \textbf{R}$_{coll}$ depending on which unit was sensitized. For instance, for the x-axis label '2' (device \textbf{M}$_2$ in Fig. \ref{fig:array-resistive-map}) four different values of \textbf{R}$_{coll}$ are displayed. They all correspond to having set \textbf{M}$_2$ as $V_t^S = 0.06$ V (top left on the map shown in Fig. \ref{fig:array-resistive-map}). The circle corresponds to the initial state $t=0$ s (as depicted in the upper panel where the external signal is displayed). The star reflects the updated value of \textbf{R}$_{coll}$ after a positive semicycle ($t=0.5$ s), the diamond is the condition after the polarity change ($t=1.0$ s), and the triangle demonstrates the state at $t=1.5$ s. The lines included in Fig. \ref{fig:Sensitization-experiment} correspond to the case of the completely homogeneous network (similar to the previous cycling experiment). The differences between the values of \textbf{R}$_{coll}$ when \textbf{M}$_2$ has been sensitized compared to the homogeneous case are notorious mainly at $t = 1$ s and $t = 1.5$ s. Moreover, the difference between the homogeneous case and the sensitized one depends on the position of the affected unit. The difference between the homogeneous and non-homogeneous (sensitized) cases remains almost unnoticed in the case of sensitizing device \textbf{M}$_4$. The maps for each remnant condition of the 24 simulation runs are included in the Supplementary Material.             

The presence of a sensitized unit - understood as a device with a lower switching threshold than the surrounding - determines the appearance of a distinctive global state for each remnant condition only when the role that specific unit displays within the array is relevant. It is worth noticing the networks display two different evolution modes. On the one hand, as it was demonstrated by Fig. \ref{fig:global-resistive-state}, upon cycling \textbf{R}$_{coll}$ switches between extreme values determined by the previous history of applied stimuli. The initial state, in which every device was configured as $R_{OFF}$, is never recovered and the highly resistance condition differs among cycles. This is also visible in the sensitized case since there are three different \textbf{R}$_{coll}$ values for three expected high resistance conditions: initially ($t=0$ s), after the first positive semi cycle ($t=0.5$ s) - RESET-like according to eq. \ref{eq:dX-dt-Pershin-f} - and after the third semi cycle $t=1.5$ s (second RESET-like condition). Additionally, there are also differences among the devices in the sensitized case - and how dramatically the sensitized unit impacts on the overall response - depends on its position within the network. Moreover, the difference between the collective state of the  undistorted network and the affected one depends on how different $V_t^S$ is compared to $V_t$, being more pronounced when $\frac{V_t}{V_t^S} = 10$ than $\frac{V_t}{V_t^S} = 5$ or $\frac{V_t}{V_t^S} = 1.2$ cases (not shown). 

The described picture is no other than the direct consequence of the voltage drop distribution. However, the conclusion can be generalized as follows: such as memristive devices' states are conditioned by the history of applied stimuli, the networks' state is a function of both the applied stimuli but also the spatial distribution of them. The preferential paths for the current to flow not only depend upon the sourced signal details but also on the choice of where to feed it into the array and the resistance distribution that such a voltage drop determines with cycling. Moreover, this spatial or lateral preference within the network affects the impact that intentional sensitization may have in the overall response of the system. This, in turn, has implications in the design of experimental strategies to drive the response of such complex systems. Taking self-assemblies as desirable experimental realizations of neuromorphic systems, and given their complexity, ideas such as intentionally modifying localized zones are supposed to help driving the response of it promoting the formation of preferential hubs. This would enable small world-like networks claimed to be promising in the context of neuromorphic implementations\cite{loeffler_topological_2020}. What we have just shown with this 4x4 simple model is that the impact of sensitizing a unit within an assembly is not independent on the pre-existent connectivity and nature of the memristive components composing the system. Further studies are required to clarify subtle aspects such as to which extent the observed phenomenology rely on the type of memristive switch, its underlying physical mechanism, or if memristive behaviour is even necessary at all (considering that progressive irreversible breakdown might display similar phenomenology) to appreciate this emergent texture and its plausibly associated usefulness in technological applications.

\section*{Supplementary Material}

\noindent This complementary document includes the maps of the internal resistance state distribution corresponding to the cycling and sensitization experiments (related to Figs. \ref{fig:global-resistive-state} \ref{fig:Sensitization-experiment}). Each map codifies in a snapshot the distribution of each unit's internal resistance state calculated at each remnant condition of the externally applied signal. Associated animations are also on-line available as multimedia views.     

\begin{acknowledgments}
\noindent C.P. Quinteros gratefully acknowledges financial support from EU-H2020-RISE project \textit{Memristive and multiferroic materials for logic units in nanoelectronics} 'MELON' (SEP-2106565560).  
\end{acknowledgments}

\section*{Data Availability Statement}

\noindent The data that support the findings of this study are openly available. Simulation of individual units has been performed using Pershin's model included by default in NGSPICE and setting the parameters detailed in the manuscript. The code to set the array is available at \href{https://github.com/fabriziodifran/mem_net}{https://github.com/fabriziodifran/MemNet}.

\bibliography{LibJAug2nd}

\end{document}